\documentclass[superscriptaddress]{revtex4}

\usepackage{color,psfig,graphics,psfrag,amsmath,amssymb,amsfonts,rotating}

\newcommand{\eps}{\varepsilon}

\begin{document}

\title{The theoretical capacity of the Parity Source Coder}

\author{Stefano Ciliberti}

\affiliation{Laboratoire de Physique Th\'eorique et Mod\`eles Statistiques,
Universit\'e de Paris-Sud, b\^atiment 100, 91405, Orsay Cedex, France}

\author{Marc M\'ezard}

\affiliation{Laboratoire de Physique Th\'eorique et Mod\`eles Statistiques,
Universit\'e de Paris-Sud, b\^atiment 100, 91405, Orsay Cedex, France}


\begin{abstract}
The Parity Source Coder is a protocol for data compression which is based on
a set of parity checks organized in a sparse random network.  We consider
here the case of memoryless unbiased binary sources.  We show that the
theoretical capacity saturate the Shannon limit at large $K$. We also find
that the first corrections to the leading behavior are exponentially small,
so that the behavior at finite $K$ is very close to the optimal one.
\end{abstract}

\pacs{ }

\maketitle

\section{Introduction}

The Parity Source Coder (PSC) is a new scheme for lossy data compression,
which uses a kind of dual approach~\cite{yedidia} to the LDPC codes used in
channel coding~\cite{mackay}.  It has been introduced
in~\cite{murayamaokada}, and discussed recently in~\cite{cimeze}
and~\cite{WaiMan}. We discuss here its theoretical performances.

\begin{figure}
  \psfrag{y1}[][][2]{$y_1$}
  \psfrag{y2}[][][2]{$y_2$}
  \psfrag{y3}[][][2]{$y_3$}
  \psfrag{y4}[][][2]{$y_4$}
  \psfrag{x1}[][][2]{$x_1$}
  \psfrag{x2}[][][2]{$x_2$}
  \psfrag{x3}[][][2]{$x_3$}
  \psfrag{x4}[][][2]{$x_4$}
  \psfrag{x5}[][][2]{$x_5$}
  \psfrag{x6}[][][2]{$x_6$}
  \psfrag{x7}[][][2]{$x_7$}
  \begin{center}
    \includegraphics[angle=-0,width=.6\textwidth]{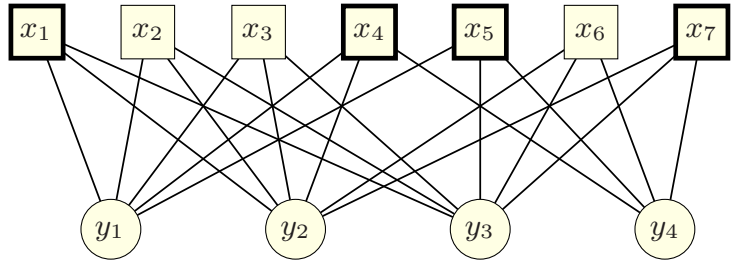}
  \end{center}
  \caption{A Tanner graph for a PSC with $M=7$ checks and $N=4$
    variables. In this example the string to be compressed is
    $\{x_1,x_2,\ldots x_7\} \equiv {\tt 1001101}$. The constraints
    $x_1,x_4,x_5,x_7$ impose the sum of the variables $y_i$ involved
    in each constraint to be $1$ mod $2$, while $x_2,x_3,x_6$ require
    that the variables add up to $0$ mod $2$. }
\label{fig:graph}
\end{figure}

\begin{figure}
  \begin{center}
    \includegraphics[angle=-90,width=.6\textwidth]{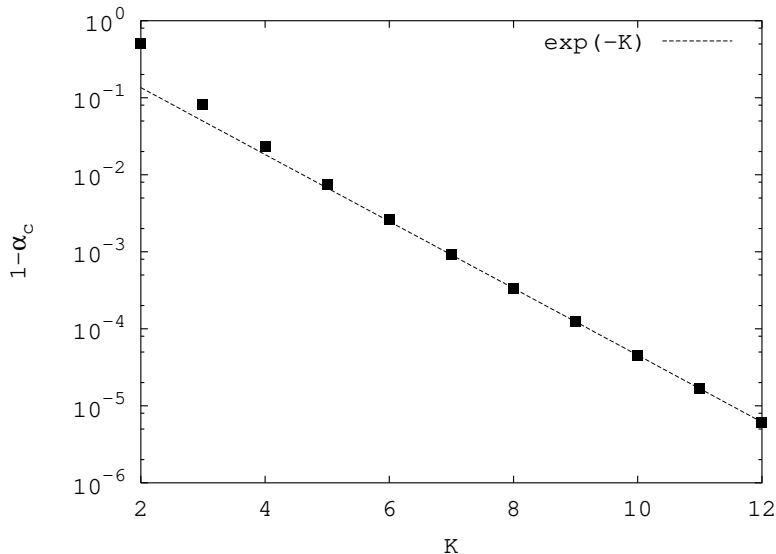}
  \end{center}
  \caption{The critical value of the control parameter marking the
    transition between SAT/UNSAT is plotted versus $K$.
    Following~\cite{merize}, one can show that the leading behavior at large
    $K$ is $\alpha_c(K)= 1 -e^{-K} - (K^2-K/2)e^{-2K} +
    \mathcal{O}(K^5e^{-3K})$ .}
\label{fig:alpha}
\end{figure}

The idea of the PSC is to use the $M$ bits ${\bf x}_M\equiv \{x_1,\ldots
x_M\}$ that we want to compress to build $M$ parity-checks on a low-density
graph involving $N (<M)$ boolean variables ${\bf y}_N\equiv \{y_1,\ldots
y_N\}$. From the theoretical point of view we will be interested in the
`thermodynamic' limit where $N$ and $M$ go to infinity while the {\em rate}
$R\equiv N/M$ is kept fixed. The topology is defined as follows: Each
constraint is connected to exactly $K$ variables chosen at random. This
implies that the probability distribution of the variable connectivity is
Poissonian (as in Erd\"os-Renyi random graphs) with mean $K\alpha$. This is
the general setting for a number of constraint satisfaction
problems~\cite{factor}. In our case such a graph (cfr. Fig.~\ref{fig:graph})
defines a set of $M$ linear equations for the $N$ variables:
\begin{eqnarray}
  y_{i_1^a} + y_{i_2^a} + \ldots + y_{i_K^a} 
  & = & x_a\,\ \text{mod}\, 2\ ,
  \quad \quad a=1,\ldots M, 
  \label{eq:linear}  
\end{eqnarray}
where $x_i,y_i\in \{0,1\}$, and the indices $i_1^a,i_2^a, \dots ,i_K^a $are
chosen in $\{1,\dots,N\}$ with uniform distribution (the repetition of two
indices in the same constraint can be forbidden, but this is irrelevant in
the large $N$ limit which interests us here). This problem is called
$K$-XORSAT~\cite{xor} and it has been recently studied in~\cite{merize}
and~\cite{cocco}. It is also a diluted version of the p-spin model used in
spin glass theory~\cite{pspin}. Here we use it to set up a data compressor,
following \cite{cimeze}. The encoded word corresponds to the solution of the
linear system~(\ref{eq:linear}) which minimizes the number of errors.  In
the thermodynamic limit, it has been shown that the critical value
$\alpha_c$ that signals the $K$-XORSAT problem has a phase transition at a
critical value $\alpha_c$ of the ratio $\alpha=M/N$. For $\alpha<\alpha_c$ a
random instance is satisfiable (in the sense that there exists an assignment
of the $N$ variables satisfying all $M$ equations) with probability
one. This is the SAT phase. For $\alpha>\alpha_c$ a random instance is
unsatisfiable with probability one: there is no assignment satisfying all
constraints.  The critical density of constraints $\alpha_c$ increases with
$K$ and goes exponentially fast to $1$ as $K$ increases
(Fig.~\ref{fig:alpha}), as can be computed using the formalism introduced
in~\cite{merize,cocco}.  The $K$-XORSAT can be used for data compression by
working in the UNSAT phase with $\alpha>1$. As the encoding step ${\bf x}_M
\to {\bf y}_N$ consists in finding the string ${\bf y}_N$ which violates the
smallest number of constraints in (\ref{eq:linear}), the compression rate is
$R=1/\alpha$.  Once we have the encoded word, the decompression step ${\bf
y}_N \to {\bf x}_M^*$ is done by setting $x^*_a=0$ or $1$ according to
eq.~(\ref{eq:linear}). The {\em distortion} is defined as the number of bits
which are not properly recovered, divided by the total number of bits
$M$. We can look at the problem in terms of a ``cost'' function
$\eps_a(y_{i_1^a} \ldots y_{i_K^a} | x_a)$ which is $0$ if
eq.~(\ref{eq:linear}) is verified and $2$ otherwise. The total cost $E$ of
the compression process is then twice the total number of unsatisfied
equations in the linear system (\ref{eq:linear}). The distortion is
related to it by
\begin{equation}
 D=\frac E {2M} = \frac{E}{2N\alpha} \ .
 \label{eq:distortion}
\end{equation}

We consider here the simplest version of the lossy compression problem: We
deal with uncorrelated unbiased binary sources, {\sl i.e.}  prob($x_1,\ldots
x_M$)=$\prod_{a=1,M}$prob($x_a$) and prob($x_a$=$0$)=prob($x_a$=$1$)=1/2.
The rate distortion theorem~\cite{coverthomas} states that a distortion $D$
can be achieved if and only if the rate is large enough, $R\geq R^*$, where
the Shannon bound $R^*$ is given by
\begin{equation*}
 R^*= 1-H_2(D)\ ,
\end{equation*}
and $H_2(x)=-x\log x-(1-x)\log(1-x)$ is the binary entropy.  Basically the
proof of achievability in this theorem relies on a choice of codewords (the
set of all possible encoded words) which is a random set.  This is
intimately related to the random energy model (REM)~\cite{derrida}.  On the
other hand, our PSC can be argued to become a random energy model in the
large $K$ limit, in the same way as the $p$-spin models becomes a REM in the
large $p$ limit~\cite{derrida,olivier}. Seen from this point of view, it is
not surprising that the performances of the PSC converge to the Shannon
bound in the limit of large $K$, as we shall prove here. In fact the same
optimal performance has been found in a recent work~\cite{perceptron} using
a a non-monotonic perceptron. Again in such a device each bit of the decoded
word is chosen to be a function of the complete encoded word, which is the
same as letting $K=N$, {\em i.e.}  infinity in the thermodynamic limit, in
our language.

However all these ``optimal'' source coding devices, based either on a
random codebook like in the REM, on a fully connected perceptron, or on the
PSC at $K\to\infty$, have a serious drawback: there is no known fast
algorithm to perform the encoding.  Physically, the encoding step is a
search of the ground state, the one which minimizes the number of violated
constraints. This has to take place in the UNSAT phase $\alpha>1$ where
these systems are frustrated. Finding the exact ground state is an
NP-complete problem, but it turns out that we don't even have good
heuristics to find approximate ground states.  Such a heuristic of course
cannot exist for the REM, but one could hope to find one for the PSC with
finite $K$. For instance in the related problem of
$K$-satisfiability~\cite{meze}, or source coding devices based on random
nodes~\cite{cimeze}, there exist good heuristics based on the message
passing ``survey propagation'' (SP) algorithm which can be seen as a
generalization of the celebrated `belief propagation'
algorithm~\cite{bmz,bz}.  While this algorithm, as such, does not work for
the PSC, it seems possible that one could develop powerful algorithms for
the finite-$K$ PSC in the future.  Actually, a very recent work~\cite{mmw}
proposes a message passing algorithm, inspired by SP, which seems to show
very good performance.  This motivates the present study of the theoretical
capacity of the PSC at finite $K$.

In this note we compute explicitely the distortion of the PSC in the limit
where the clause connectivity $K$ becomes large. We first show that for
$K\to\infty$ the distortion becomes optimal (it saturates the Shannon
bound).  As for the finite $K$ corrections, we find that, for a given value
of the rate $R=1/\alpha$, the distortion is
\begin{equation}
  D = D_{Sh} + a\sqrt{K} e^{-K\Delta}\left(1+\mathcal{O}(1/K)\right) \ ,
\end{equation}
where $D_{Sh}$ satisfies $1-H_2(D_{Sh})= 1/\alpha$ and the coefficients $a$
and $\Delta$ depend on $\alpha$. In particular, the actual $\Delta$ lies in
$[\log 2, 1]$, and goes to $\log 2$ in the large $K$ limit.  The fact that
the first finite-$K$ corrections are exponentially small must be stressed:
This means that also a parity source coder with $K=5$ or $6$ is in practice
nearly optimal. A good encoding algorithm for this case could thus turn this
PSC into a very good compressor.  We stress that the range of validity of
the result of this paper is limited to the case of uncorrelated
sources. This is confirmed by the statistical description of a family of
code ensembles presented in~\cite{hosakaba}. On the other hand, the
hypothesis of a non-biased input message does not seem to play a role.

As we mentioned previously, a protocol very similar to this PSC (the only
difference being the underlying graph topology) has been introduced
in~\cite{murayamaokada}, and Murayama~\cite{murayama} has shown that some
belief-propagation based algorithm can be used for encoding in the $K=2$
case.  Our result shows that the optimal capacity ({\it i.e.} Shannon's
bound) can be obtained only in the limit of large $K$, at variance with some
of the statements in~\cite{murayamaokada}.  It gives the analog, for source
coding, to the result of Kabashima and Saad~\cite{kabasaad99} on channel
capacity of error-correcting codes at large $K$.

\section{Cavity equations}

In order to deal with the $K$-XORSAT problem we take advantage of the cavity
method as explained in \cite{meze}. This method is heuristic (the main
assumptions that can be checked self-consistently) but it is believed to be
exact. As for the $K$-XORSAT problem, its range of validity has been
rigorously established in \cite{merize} and \cite{moripa}. In particular, the
cavity result for the critical threshold $\alpha_c$ is exact. For $\alpha >
\alpha_c$ (the regime where we use it) this method finds the correct
ground-state energy up to a threshold value $\alpha_G$, which is $\simeq
3.07$ for $K=3$~\cite{moripa} and increases with $K$ as one can see from
numerics. 

\begin{figure}
  \psfrag{ph0}[][][1.5]{$h_{1\to a}$}
  \psfrag{ph1}[][][1.5]{$h_{2\to a}\ \ $}
  \psfrag{ph2}[][][1.5]{$h_{3\to a}$}
  \psfrag{ph3}[][][1.5]{$h_{4\to a}\ \ \ $}
  \psfrag{uai}[][][1.5]{$u_{a\to 0}\ \ \ \ $}
  \psfrag{4}[][][1.5]{$a$}
  \psfrag{5}[][][1.5]{$0$}
  \psfrag{0}[][][1.5]{$1$}
  \psfrag{1}[][][1.5]{$2$}
  \psfrag{2}[][][1.5]{$3$}
  \psfrag{3}[][][1.5]{$4$}
  \psfrag{6}[][][1.5]{$$}
  \psfrag{7}[][][1.5]{$$}
  \psfrag{8}[][][1.5]{$$}
  \psfrag{9}[][][1.5]{$$}
  \psfrag{10}[][][1.5]{$$}
  \psfrag{11}[][][1.5]{$$}
  \psfrag{12}[][][1.5]{$$}
  \psfrag{13}[][][1.5]{$$}
  \psfrag{14}[][][1.5]{$$}
  \psfrag{15}[][][1.5]{$$}
  \psfrag{16}[][][1.5]{$$}
  \psfrag{17}[][][1.5]{$$}
  \psfrag{18}[][][1.5]{$$}
  \begin{center}
    \includegraphics[angle=-0,width=.8\textwidth]{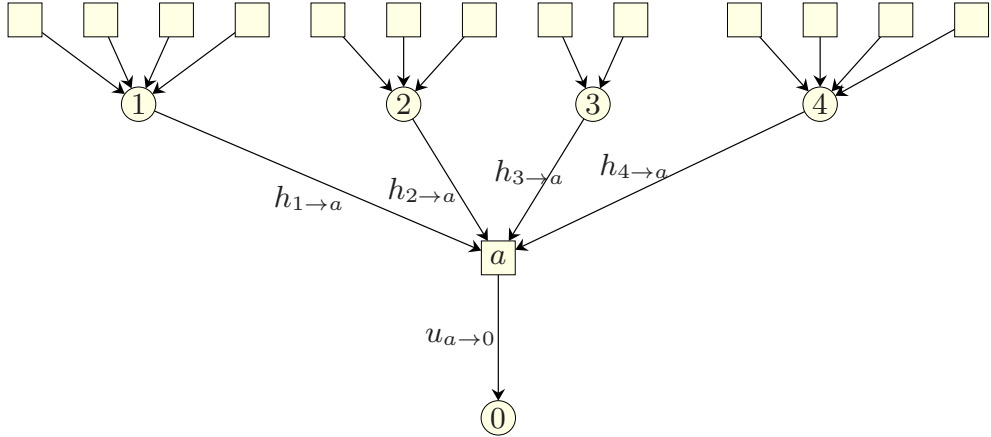}
  \end{center}
  \caption{The iterative idea behind the cavity equations is illustrated
    here for $K=5$.}
\label{fig:recursion}
\end{figure}

For the sake of simplicity, we pass from boolean variables to Ising spins,
thus taking values in $\{-1,+1\}$. The general idea behind the cavity
approach is summarized in Fig.~\ref{fig:recursion}. Since the local
structure of the random graph is tree-like, we focus on a single clause and
look at the variables connected to it. We introduce two types of messages,
cavity biases $u_{a\to i}$ going from clause $a$ to variable $i$, and cavity
fields $h_{i\to a}$ going from variable $i$ to clause $a$. A cavity bias can
be $0$ (which means that, as for the clause $a$, variable $i$ is free to
assume any value), or $\pm 1$ (meaning that this is the value that $i$
should take in order to satisfy clause $a$). The message sent from clause
$a$ must take into account all the other variables connected to it; each of
these sends to $a$ a cavity field which is nothing but the sum of all the
other incoming cavity biases:  $h_{j\to a} = \sum_{b\in j - a} u_{b\to j}$.
In the most general case, the space of low-energy configurations is broken
into many disconnected components (clustering). The general object we need
to deal with this is then a functional distribution $\mathcal{Q}[{\sf
q}(u)]$ giving the probability that, if one link $a\to i$ is chosen at
random, the probability (with respect to the choice of the cluster) of
observing a bias $u_{a\to i}$ is ${\sf q}_{a\to i}(u_{a\to i})$. The same
holds for the distribution of cavity fields, $\mathcal{P}[{\sf p}(h)]$.  We
thus suppose to have a population of ${\sf q}(u)$'s and ${\sf p}(h)$'s. In
order to simplify the notations, we shall simply call $u_0$ the bias on
variable $0$, with no regards about the clause it is coming from. According
to~\cite{merize}, we iterate the following self-consistent equations:
\begin{eqnarray}
  {\sf q}_{0}(u_{0}) 
  &=& \sum_{h_{1},\ldots h_{{(K-1)}}} 
  {\sf p}^{(p_1)}_{1}(h_{1})
  \cdots {\sf p}^{(p_{K-1})}_{(K-1)}(h_{(K-1)}) 
  \delta\bigg(u,S(Jh_{1}\cdots
  h_{(K-1)})\bigg) \ , \quad \textrm{with prob.} 
  \quad \prod_{i=1}^{K-1} f_{K\alpha}(p_i)
  \label{eq:qu}\\
  {\sf p}^{(p)}(h) &=& \frac 1{A^{(p)}(y)} \sum_{u_1,\ldots u_p} 
  {\sf q_1}(u_1)\cdots {\sf q_p}(u_p) 
  \delta\left(h,\sum_{a=1}^p u_a\right) 
  \exp \bigg\{y \Big|\sum_{a=1}^p u_a \Big| -  y\sum_{a=1}^p  |u_a|\bigg\} \
  , \label{eq:ph}
  \\
  A^{(p)}(y) &=& \sum_{u_1,\ldots u_p} {\sf q_1}(u_1) 
  \cdots {\sf q_p} (u_p) \exp\bigg\{y\Big|\sum_{a=1}^p u_a 
  \Big|-y\sum_{a=1}^p|u_a|\bigg\}  \ . \label{eq:A}
\end{eqnarray}
Here $S(x)\equiv$ sign$(x)$ for $x\neq 0$, $S(0)\equiv 0$, and
$f_{K\alpha}(\cdot)$ is the Poisson distribution with mean $K\alpha$.  The
first of these equations is the direct implementation of the recursion
illustrated in Fig.~\ref{fig:recursion}: The delta function ensures that
clause $a$ sends the proper value to variable $0$. In the second equation, a
reweighting term is present~\cite{meze}. This is due to the fact that if we
add one variable and want to compute the new probability distributions at a
given value of the energy $E$, then we need all the contributions from the
states at energy $E-\Delta E$, where $\Delta E$ is the energy shift caused
by the addition of one variable. If the number of clusters at energy $E$ is
exp$(N\Sigma(E/N))$, then the expansion $\Sigma(E)\simeq\Sigma(E-\Delta E) -
y \Delta E$ leads to a reweighting exp$(-y\Delta E)$, with
$y=\partial\Sigma/\partial E$.  The knowledge of these distributions allows
to compute the free energy $\Phi(y)$:
\begin{eqnarray}
  \label{eq:phi}
  \Phi(y) &=& \Phi_1(y) - (K-1)\alpha\Phi_2(y)\ , \\
  \Phi_1(y) &=& -\frac 1y \overline{\log A^{(p)}(y)}\ ,\label{eq:phi1}\\  
  \Phi_2(y) &=& -\frac 1 y \overline{\log\sum_{u} {\sf q}(u;\{p_i\}) 
    \sum_h {\sf p}^{(p)}(h) e^{y(|u+h|-|u|-|h|)}} \ , \label{eq:phi2}
\end{eqnarray}
where the average is taken over the random graph ensemble and over the
population of the distributions ${\sf q}(u)$'s and ${\sf p}(h)$'s. The free
energy in (\ref{eq:phi}) is obtained by adding one variable (and a certain
number of clauses) to a system with $N$ variables and computing the
contribution arising from the corresponding shift in energy,
exp$(-y\Phi_1)=\langle$exp$(-y\Delta E)\rangle$. The correction term is due
to the fact that in the $(N+1)-$variables system the probability of
generating the clauses is slightly lower thus we have to cancel a fraction
of them at random (see \cite{meze} for a detailed derivation). The
ground-state energy is then evaluated as the min$_y \Phi(y)$.

Actually, the nature of messages allow for a simplification of the
cavity equations: We write
\begin{equation}
  {\sf q}(u) = \eta \delta_{u,0} + \frac{1-\eta}{2} 
  \left [\delta_{u,-1} + \delta_{u,+1}\right] \ .
  \label{eq:qudiscreto}
\end{equation}
Also, it should be clear that, as for the ${\sf p}(h)$, what matters is only
the sign of the field $h$, then:
\begin{equation}
  {\sf p}^{(p)}(h) = 
  \frac 1 {A^{(p)}} \left( w_0^{(p)} \delta_{S(h),0} + 
  w_+^{(p)} \delta_{S(h),+1} + w^{(p)}_-\delta_{S(h),-1}
  \right) \ ,
  \label{eq:phdiscreto}
\end{equation}
with $A=w_0+w_++w_-$ and $w_+=w_-$ because of the up-down symmetry of
the problem. In practice, one needs to work with a single population
of real numbers $\eta_i$, that leads to a stationary distribution
$\rho(\eta)$.  For any fixed value of $y$, the self-consistent
equations (\ref{eq:qu}, \ref{eq:ph}, \ref{eq:A}) are solved as
follows:
\begin{enumerate}
\item Consider a population of $\eta_i$ randomly distributed in
  $[0,1]$. 
  \item Do $K-1$ times:
    \begin{itemize}
    \item Pick a random integer $p$ with probability $f_{K\alpha}(p)$.
    \item Choose $p$ values $\eta_1,\ldots \eta_p$ and compute a
      probability distribution ${\sf p}(h)$ according to
      (\ref{eq:ph}). Given (\ref{eq:phdiscreto}), this amounts to
      computing two real numbers: $w_0$ and the normalization $A$.
    \item Compute $\Phi_1$ as in (\ref{eq:phi1}) through this $A$.
    \end{itemize}
  \item Using these $K-1$ distributions ${\sf p}(h)$'s, compute a new ${\sf
    q}(u)$ according to (\ref{eq:qu}). Given (\ref{eq:qudiscreto}) this is
    the same as computing a new value $\eta_0$.
  \item Use this new ${\sf q}(u)$ and a new extracted ${\sf p}(h)$ to
  compute $\Phi_2$ as in (\ref{eq:phi2}). The total free energy can now be
  evaluated via (\ref{eq:phi}).
  \item Replace an $\eta$ value randomly chosen in the distribtion
  with the new value $\eta_0$.
  \item Go to step {\tt 2} until a stationary distribution $\rho(\eta)$ is
  reached. (The free energy attains then a stationary value.)
\end{enumerate} 

We are now going to discuss the cavity equations for large $K$ and we will
use the algorithm we have just described to check numerically our asymptotic
results.

\section{The Shannon bound}

The cavity equations (\ref{eq:qu}, \ref{eq:ph}, \ref{eq:A}) have been
discussed in~\cite{merize} mainly concerning the value of $\alpha_c(K)$ and
the behavior of the ground state energy $E_0(K)$ close to $\alpha_c(K)$. We
want to compute $E_0(K)$ at any $\alpha$ in the large-$K$ limit.

For large $K$, there is a self-consistent solution of the cavity equations
such that all the $w_0$ are very small, in fact exponentially small.  We
just need to assume that the typical value of a $w_0$ is much smaller than
$1/K$.  This condition on $w_0$'s shows that $\eta$ is zero to leading
order, because from eq.~(\ref{eq:qu}) one finds that
\begin{equation}
  \label{eq:eta}
  \eta=1-\prod_{i=1}^{K-1} \left(1-w_0^{(p_i)}\right) \ .
\end{equation}
We shall be more precise below as we verify self-consistently the
assumption on $w_0$ and will be able to compute the first non-zero
term. Here we work directly with $\eta=0$.  We need to compute the new
value of $w_0$ and $w_+$ using eq.~(\ref{eq:ph}). If $K$ is large, $p$
is generically large (it is Poisson distributed with mean $K \alpha$).
If $p$ is even (the case of $p$ odd is an immediate generalization)
one finds:
\begin{eqnarray}
  w_0^{(p)} &=& \binom{p}{p/2} \frac{e^{-py}}{2^p} \simeq 
  \frac{2 e^{-py}}{\sqrt{2\pi p}} \ , \\
  w_+^{(p)} &=& \frac 1{2^p} 
  \sum_{q=0}^{\frac p 2 -1}
  \binom{p}{q} e^{-qy} \simeq \frac p{2^p} 
  \int_0^{1/2} \frac {dx}{\sqrt{2\pi p x(1-x)}} 
  \exp\Big\{p\left(-x\log x -(1-x) \log(1-x) -
        xy\right) \Big\} \ ,\\
  w_-^{(p)} &=& w_+^{(p)} \ .
\end{eqnarray}
The integral can be evaluated for $p$ large by the saddle point method (the
saddle point being $x^*=-y+\log(1+e^y)$) and we have
\begin{equation*}
  w_+^{(p)} = \left(\frac{1+e^{-2y}}{2}\right)^p \ . 
\end{equation*}
Since for any finite $y$ this is exponentially larger than $w_0$, the
leading term in the normalization constant is just
\begin{equation}
  A^{(p)}(y) = 2w_+^{(p)}= 2 \left(\frac{1+e^{-2y}}{2}\right)^p \ . 
  \label{eq:Ainf}
\end{equation}
Now, it is not difficult to show that eq.~(\ref{eq:phi2}) can be rewritten
as
\begin{equation}
  \Phi_2(y) = -\frac 1y \overline{
    \log\left(\frac{A^{(p+1)}(y)}{A^{(p)}(y)}\right)} \ ,
\end{equation}
and thus the free-energy can be computed from the
normalization~(\ref{eq:Ainf}) alone. We find that
\begin{equation}
  \Phi(y) = -\frac 1y \left(\overline{ \log A^{(p)}(y) } 
  -(K-1)\alpha \overline{\log\left(\frac{A^{(p+1)}(y)}{A^{(p)}(y)}\right)}
  \right) = -\frac 1y \left[ 
  \log 2 + \alpha \log\left(\frac{1+e^{-2y}}{2}\right) \right] 
  \equiv \Phi_{\infty}(y) \ . 
\end{equation}
The ground state energy is the maximum of $\Phi(y)$~\cite{meze} and,
according to eq.~(\ref{eq:distortion}), this gives a distortion $D$ for the
parity source coder at large $K$
\begin{equation}
  D=\frac{1}{2 \alpha} \max_y \Phi_\infty(y) \ .
  \label{eq:dres}
\end{equation}
The Shannon bound says that the minimum distortion satisfies
$1-H_2(D_{Sh})=1/\alpha$. A few lines of computation show that the distortion
in~(\ref{eq:dres}) actually saturates the Shannon bound. Let's call $z$ the
value of $y$ where $\Phi_{\infty}(y)$ is maximal. It satisfies:
\begin{equation}
  \label{eq:yinf}
  \log 2 = \alpha \left(z \tanh z - \log \cosh z\right) 
\end{equation}
Then one gets
\begin{equation}
  \frac{\Phi_\infty(z)}{2 \alpha}= \frac{1}{e^{2z}+1}
  \quad \Rightarrow\quad
  H_2\left(\frac{\Phi_\infty(z)}{2 \alpha} \right)=
  \frac{1}{\log 2}\left(-\log \frac{1}{e^{2z}+1}
  -2 z  \frac{e^{2z}}{e^{2z}+1}\right) \ .
  \label{eq:hinf}
\end{equation}
After some algebra one can derive from this the seeked result:
\begin{equation}
  H_2\left(\frac{\Phi_\infty(z)}{2 \alpha} \right)=1 - \frac{1}{\alpha} \ . 
\end{equation}
This shows that at very large $K$ the XORSAT problem gives exactly the
Shannon limit. We now look at finite-$K$ corrections in order to see how
this asymptotic performance is reached.

\section{Corrections}

In order to compute the first order corrections to the leading
behavior we compute the normalization constant in~(\ref{eq:A}) under
the hypothesis of small (but finite) $\eta$:
\begin{eqnarray}
  A^{(p)}(y) &=& \prod_{a=1}^p \left(\frac{1-\eta_a}{2}\right) 
  e^{-py} \sum_{q=0}^p  \binom{p}{q} e^{y|p-2q|} + 
  p\eta_1 \prod_{a=2}^p
  \left(\frac{1-\eta_a}{2}\right) e^{-(p-1)y} \sum_{q=0}^{p-1}
  \binom{p-1}{q} e^{y|p-1-2q|} + \ldots  
  \nonumber\\  
  &=& \frac{e^{-py}}{2^p} g_p(y) (1-\sum_{a=1}^p \eta_a+\mathcal{O}(p
  \eta)^2) + p\eta \frac{e^{-(p-1)y}}{2^{p-1}} g_{p-1}(y)
  (1+\mathcal{O}(p\eta)) + 
  \frac{e^{-(p-2)y}}{2^{p-2}} g_{p-2}(y) \mathcal{O}(p\eta)^2 + \ldots \ ,
  \label{eq:Asviluppo}\\
  g_p(y)& \equiv &\sum_{q=0}^p \binom{p}{q} e^{y|p-2q|} \ .
\end{eqnarray}
As we have shown above, the whole free energy can be computed from the
knowledge of $A^{(p)}(y)$. In order to calculate it, we compute the
function $g_p(y)$ in the large $p$ limit. We first notice that it can
be written as
\begin{equation}
  g_p(y) = \sum_{\sigma_1,\ldots \sigma_p} \exp\left[y\Big|\sum_{i=1}^p
  \sigma_i\Big|\right] \ ,
\end{equation}
where $\sigma_i$ are Ising spins. Thus, 
\begin{equation}
  g_p(y) + g_p(-y) = \sum_{\{\sigma_i\}} \left[e^{y|\sum_i \sigma_i |} +
    e^{-y|\sum_i \sigma_i |}\right] = \sum_{\{\sigma_i\}} \left[e^{y \sum_i
      \sigma_i } + e^{-y\sum_i \sigma_i }\right] = 2 (2\cosh y)^p\ .
  \label{eq:gpp}
\end{equation}
We use a Fourier transformation to express $g_p(-y)$:
\begin{eqnarray*}
  g_p(-y) &=& \frac y\pi \int \frac {dk}{k^2+y^2} e^{ikp} \sum_{q=0}^p
  \binom{p}{q} e^{-i2qk} 
  =\frac y\pi 2^p \int \frac {dk (\cos k)^p}{k^2+y^2} 
  = \frac y\pi \ 2^p \sqrt{\frac{2\pi}{p}}
  \left(1-\frac 1{4p}+\mathcal{O}\left(\frac 1{p^2}\right)\right)
  \sum_{n=-\infty}^{+\infty} \frac {(-1)^{np}}{( \pi n)^2 +y^2} \ ,  
\end{eqnarray*}
The sum can be done exactly and we have
\begin{equation}
  g_{p\ even}(-y) = \frac{2^{p+1}}{\sqrt{2\pi p}}\frac 1{\tanh
  y}(1-1/4p+\mathcal{O}(1/p^2)) \ , \qquad 
  g_{p\ odd}(-y) =
  \frac{2^{p+1}}{\sqrt{2\pi p}}\frac 1{ \sinh (y)} 
  (1-1/4p+\mathcal{O}(1/p^2)) \ .
\end{equation}
Using (\ref{eq:gpp}), we get for $p$ even
\begin{equation}
  g_p(y) = 2^{p+1} (\cosh y)^p \left[1-\frac{(\cosh y)^{-p}}{\sqrt{2\pi p} 
    (\tanh y ) } \left(1-\frac 1{4p}+\mathcal{O}\left(\frac{1}{p^2}\right)
  \right)\right]\ ,
\end{equation}
with the replacement $\tanh y$ $\to$ $\sinh y$ if $p$ is odd. To the
leading order we have thus 
\begin{equation}
  A^{(p)}(y) = 2 \left(\frac {1+e^{-2y}}{2}\right)^p
  \left(1+\mathcal{O}( p^{\gamma} e^{-p} )\right) \ ,
\end{equation}
with some exponent $\gamma$ which depends on the actual order of magnitude
of $\eta$. To compute it we first need to know the weight for $h=0$. If $p$
is even we use eq.~(\ref{eq:ph}) and we note that the main contribution (in
the same hypothesis of $\eta$ small) is given by
\begin{eqnarray}
  w_0^{(p\ even)} &=& \frac 1{A^{(p)}(y)}e^{-py} \binom{p}{p/2} \frac 1{2^p}
  \prod_{a=1}^p (1-\eta_a) +\mathcal{O}(p\eta) \nonumber \\ 
  &=& \frac 12 \left(\frac {1+e^{-2y}}{2}\right)^{-p} 
  \left(1+\mathcal{O}( p^{\gamma} e^{-p} )\right) e^{-py}
  \frac{2^{p+1}}{\sqrt{2\pi p}}\left(1-\frac 1{4p} + 
  \mathcal{O}\left(\frac 1{p^2}\right)\right)
  \frac 1{2^p}(1+\mathcal{O}(p\eta)) \nonumber \\ 
  &=& \frac{(\cosh y)^{-p}}{\sqrt{2\pi p}}
  \left( 1 -1/4p + \mathcal{O}(1/p^2) +\mathcal{O}(p\eta)   \right)   \ .
  \label{eq:wopeven}
\end{eqnarray}
(Here we have also assumed that $p>0$, since $w_0=1$ if $p=0$.) 
On the other hand, if $p$ is odd we have
\begin{eqnarray*}
  w_0^{(p\ odd)}&=& \eta \frac 1{A^{(p)}(y)}e^{-(p-1) y} p
  \binom{p-1}{(p-1)/2} \frac 1{2^{p-1}} \prod_{a=1}^{p-1} (1-\eta_a)
  +\mathcal{O}(p\eta)^2 \\ 
  &=& \sqrt{\frac{p}{2\pi}} \eta e^y (\cosh y)^{-p} (1+\mathcal{O}(1/p)) \ .
\end{eqnarray*}
To the leading order, $\eta$ does not fluctuate and takes the value
\begin{eqnarray*}
  \eta \simeq -\log(1-\eta) \simeq \sum_{i=1}^{K-1} w_0^{(p_i)} 
  &=& (K-1) \left[
    e^{-K\alpha} + e^{-K\alpha}( \cosh(K\alpha) -1) 
    \overline{w_0^{(p\ even)}}\big|_{p>0} + 
    e^{-K\alpha}\sinh(K\alpha)     \overline{w_0^{(p\ odd)}} 
    \right] \\  
  & \simeq & 
  \frac {K-1}{2}\overline{\left[\frac{(\cosh y)^{-p}}{\sqrt{2\pi p}}
      \left(1-\frac 1{4p}+
      \mathcal{O}\left(\frac 1{p^2}\right)\right)\right]}_{p\ even >0}  + 
  \mathcal{O}(Ke^{-K\alpha}) + \mathcal{O}(\eta^2)\ ,
\end{eqnarray*}
since the two other terms ($p=0$ and $p$ odd) are exponentially subleading. 
In order to perform this average we use 
\begin{equation}
  \frac 1{p^z} = \frac 1{\Gamma(z)} \int dt\  t^{z-1} e^{-pt} 
\end{equation}
to express the denominator. This allows to perform the average over $p$
even. We then have
\begin{eqnarray*}
  \eta & = & \frac {Ke^{-K\alpha} }{4\pi\sqrt{2}}\left(1-\frac 1K\right)
  \left\{ 2\int dt\, t^{-1/2} (\cosh(\beta e^{-t})-1) - 
  \int dt\, t^{1/2} (\cosh(\beta e^{-t})-1) + 
  \mathcal{O}\left(\beta^{-5/2}\right)\right \} + \mathcal{O}(Ke^{-K\alpha})\\
  & = & \frac {Ke^{-K\alpha}} {4\pi\sqrt{2}} 4\beta
  \int dt \,t^{1/2}e^{-t} \sinh(\beta e^{-t})\left(1-\frac t6+ \mathcal{O}(t^2)
  \right)\left(1-\frac 1K\right) + \mathcal{O}(Ke^{-K\alpha})
\end{eqnarray*}
where $\beta\equiv K\alpha/\cosh y$. We then set $t=\tau/\beta$ and expand
in $1/\beta$. This gives
\begin{equation}
  \eta = \frac{\sqrt{\cosh y}}{2\sqrt{2\pi\alpha}} K^{1/2} 
  e^{-K\alpha(1-1/\cosh y)} 
  \left[1+ \frac 1K \left(\frac{\cosh y}{8\alpha}-1\right) +
  \mathcal{O}\left(\frac{\cosh y}{K \alpha}\right)^2\right] + 
  \mathcal{O}(Ke^{-K\alpha})\ ,
  \label{eq:eta}
\end{equation}
which shows a posteriori that the small $\eta$ hypothesis is consistent. 
We now go back to (\ref{eq:Asviluppo}) and get:
\begin{eqnarray}
  A^{(p)}(y) 
  &=& 2\left(\frac{1+e^{-2y}}{2}\right)^p \left[1-p \eta \tanh y
  + \mathcal{O}(p\eta)^2\right] \nonumber\\
  &=& 2\left(\frac{1+e^{-2y}}{2}\right)^p 
  \left[ 1 - \frac{p\tanh y\sqrt{\cosh y}}{2\sqrt{2\pi\alpha}} 
      K^{1/2}e^{-K\alpha(1-1/\cosh y)}
      \left(1+\frac 1K \left(\frac{\cosh y}{8\alpha}-1\right) +
      \mathcal{O}\left(\frac{\cosh y}{K \alpha}\right)^2 \right)
      \right. + \nonumber \\
  &&  \hspace{3cm} + \mathcal{O}(pKe^{-K\alpha})\bigg] \ .
\end{eqnarray}
From this result and from eq.~(\ref{eq:phi2}) one finds that
\begin{equation}
  \Phi_2(y) = -\frac 1y \left(
  \log\left(\frac{1+e^{-2y}}{2}\right) -\eta \tanh y + \mathcal{O}(K\eta)^2
  \right) \ .
  \label{eq:phi2K}
\end{equation}
Moreover, 
\begin{equation}
  \Phi_1(y) = -\frac 1y \left( \log 2 + K\alpha
  \log\left(\frac{1+e^{-2y}}{2}\right) +
  K\alpha\eta \tanh y + \mathcal{O}(K\eta)^2 \right)\ .
  \label{eq:phi1K}
\end{equation}
We can now compute the total free energy (\ref{eq:phi}). One can check
directly that the leading corrections to the infinite $K$ limit, of order
$\mathcal{O}\Big[K^{3/2}\exp\big(-K\alpha(1-1/\cosh y)\big)\Big]$, vanish. 
We are then left with 
\begin{eqnarray}
  \Phi(y)
  &=& -\frac 1y \left[ 
  \log 2 + \alpha \log\left(\frac{1+e^{-2y}}{2}\right) \right] 
  + \frac {\sqrt{\cosh y} \tanh y} {2y\sqrt{2\pi}} (\alpha K)^{1/2} 
  e^{-K\alpha(1-1/\cosh y)} \left(1+
  \frac 1K \left(\frac{\cosh y}{8\alpha}-1\right) +
  \mathcal{O}\left(\frac{1}{K^2}\right)\right) \nonumber \\
  & = &\Phi_\infty(y) + \Delta\Phi_K(y) \ ,
  \label{eq:free}
\end{eqnarray}
where $\lim_{K\to\infty} \Delta\Phi_K = 0 $. We assume that the
maximum of the $\Phi(y)$ in (\ref{eq:free}) is at $y=z+\eps$, where
$\eps$ is exponentially small at large $K$ (we shall verify
self-consistently this hypothesis) and $z$ is the solution of
eq.~(\ref{eq:yinf}). The condition $\Phi'(y)=0$ then results in
\begin{equation}
  \eps = -\frac{\Delta\Phi'_K(z)}{\Phi''_\infty(z)} = 
  \mathcal{O}\left( K^{-1/2} e^{-K\alpha(1-1/\cosh z)}\right) \ ,
\end{equation}
where the dependence of $z$ on $\alpha$ is extracted from (\ref{eq:yinf}).
One finds that $z$ is a monotonic decreasing function. In particular, $z\sim
\sqrt{2 \textrm{log} 2/\alpha}$ at large $\alpha$ while $z$ diverges as
$(-1/2)$log$(\alpha-1)$ as $\alpha\to 1$: It follows that $\eps$ is
exponentially small in any case. Coming back to eq.~(\ref{eq:dres}), it is
then easy to see that to the leading order
\begin{equation}
  D = \frac 1{2\alpha} \left( \Phi_\infty(z) + \Delta\Phi_K(z) \right) =
  D_{Sh} + C_K(\alpha) \ , 
\end{equation}
where the corrections $C_K(\alpha)$ are finally 
\begin{equation}
  C_K(\alpha) = \frac {\sqrt{\cosh z} \tanh z} {4z\sqrt{2\pi\alpha}} K^{1/2} 
  e^{-K\alpha(1-1/\cosh z)} \left(1+
  \frac 1K \left(\frac{\cosh z}{8\alpha}-1\right) +
  \mathcal{O}\left(\frac{1}{K^2}\right)\right) 
  \left(1 + \mathcal{O}\left(K^{1/2} e^{-K\alpha}\right)\right) \ ,
  \label{eq:correction}
\end{equation}
$z$ being the solution of eq.~(\ref{eq:yinf}).

\begin{figure}
  \begin{center}
    \includegraphics[angle=-90,width=.6\textwidth]{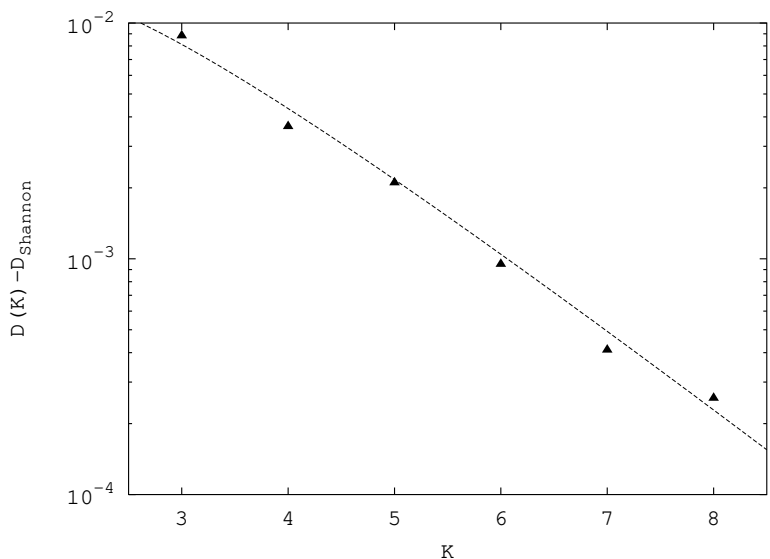}
  \end{center}
  \caption{Theoretical capacity of the PSC, $\alpha=1.3$.}
\label{fig:plot1.3}
\end{figure}

\begin{figure}
  \begin{center}
    \includegraphics[angle=-90,width=.6\textwidth]{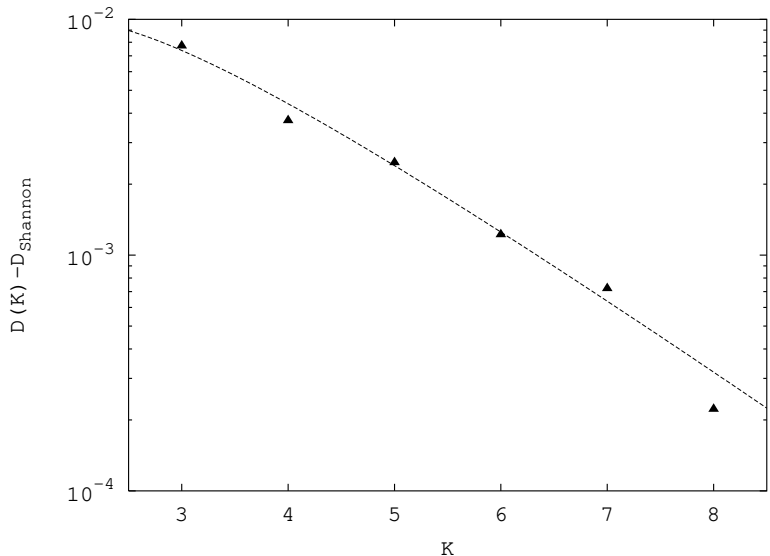}
  \end{center}
  \caption{Theoretical capacity of the PSC, $\alpha=2.0$.}
\label{fig:plot2.0}
\end{figure}

We now look at numerical data in order to verify our analytical
prediction. In Fig.~\ref{fig:plot1.3} we plot the difference between
the actual distortion of the PSC as obtained from the numerical
solution of the cavity equations at $\alpha=1.3$ and the corresponding
Shannon value. The curve is the theoretical prediction in
(\ref{eq:correction}), where we neglected the $1/K^2$ corrections. The
same plot but for $\alpha=2$ is shown in Fig.~\ref{fig:plot2.0}. In
both cases there is a very good agreement with the analytical
prediction.

\section{Conclusions}

\begin{figure}
  \begin{center}
    \includegraphics[angle=-90,width=.6\textwidth]{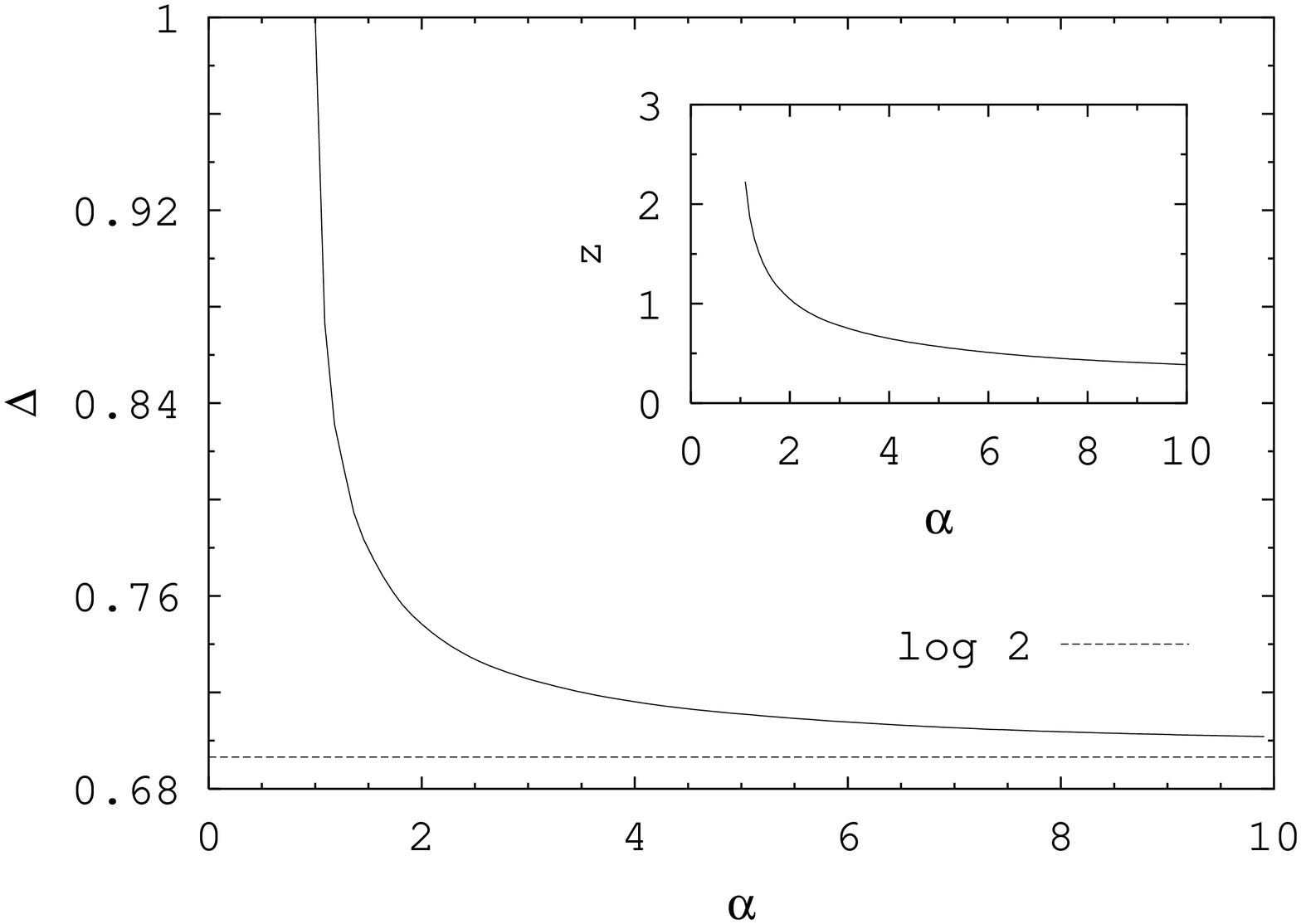}
  \end{center}
  \caption{The value of $\Delta\equiv \alpha(1-1/\cosh z)$, $z$ being the
  solution of eq.~(\ref{eq:yinf}), is plotted vs $\alpha$. One finds that
  $\Delta = \log 2+\mathcal{O}(1/\alpha)$ at large $\alpha$, while
  $\Delta=1-2\sqrt{\alpha-1}+\mathcal{O}(\alpha-1)$ as $\alpha \to
  1^+$. Inset: The actual value of $z$ as a function of $\alpha$, as given
  by eq.~(\ref{eq:yinf}). It diverges as $-\log(\alpha-1)$ as $\alpha\to
  1^+$.}
\label{fig:delta}
\end{figure}

We have shown that the theoretical capacity of the Parity Source Coder
is optimal at large $K$ and that the corrections to the leading behaviour
are exponentially small. Nevertheless, due to the smallness of $\Delta$
(cfr. Fig.~\ref{fig:delta}), the exponential decreases quite slowly, and
$1/K$ corrections are needed to take into account the deviations from the
leading behavior at relatively small values of $K$. 

\emph{Acknowledgments.} We thank O.~Rivoire M.~Wainwright, J.S.~Yedidia and
R.~Zecchina for important discussions, and an anonimous referee for useful
suggestions and references.  S.~C. is supported by EC through the network
MTR 2002-00307, DYGLAGEMEM. This work has been supported in part by the EC
through the network MTR 2002-00319 STIPCO and the FP6 IST consortium
EVERGROW.

\end{document}